# Comparison of two superconducting phases induced by a magnetic field in UTe$_2$

W. Knafo,[1,*] M. Nardone,[1] M. Valiska,[2] A. Zitouni,[1] G. Lapertot,[2] D. Aoki,[2,3] G. Knebel,[2] D. Braithwaite[2]

[1] *Laboratoire National des Champs Magnétiques Intenses, UPR 3228, CNRS-UPS-INSA-UGA, 143 Avenue de Rangueil, 31400 Toulouse, France*
[2] *Univ. Grenoble Alpes and CEA, IRIG-PHELIQS, F-38000 Grenoble, France*
[3] *Institute for Materials Research, Tohoku University, Ibaraki 311-1313, Japan*

\* Corresponding author: william.knafo@lncmi.cnrs.fr

29/06/2020 10:59

## Abstract

Superconductivity induced by a magnetic field near metamagnetism is a striking manifestation of magnetically-mediated superconducting pairing. After being observed in itinerant ferromagnets, this phenomenon was recently reported in the orthorhombic paramagnet UTe$_2$. Under a magnetic field applied along the hard magnetization axis **b**, superconductivity is reinforced on approaching metamagnetism at $\mu_0 H_m \approx 35$ T, but it abruptly disappears beyond $H_m$. On the contrary, field-induced superconductivity was reported beyond $\mu_0 H_m \approx 40$-$50$ T in a magnetic field tilted by $\simeq 25$-$30$ ° from **b** in the (**b**,**c**) plane. Here we explore the phase diagram of UTe$_2$ under these two magnetic-field directions. Zero-resistance measurements permit to confirm unambiguously that superconductivity is established beyond $H_m$ in the tilted-field direction. While superconductivity is locked exactly at fields either smaller (for a **H** || **b**), or larger (for **H** tilted by $\simeq 27$ ° from **b** to **c**), than $H_m$, the variations of the Fermi-liquid coefficient in the electrical resistivity and of the residual resistivity are surprisingly similar for the two field directions. The resemblance of the normal states for the two field directions puts constraints for theoretical models of superconductivity and implies that some subtle ingredients must be in play.

Unconventional superconductivity is observed in an ever-growing number of correlated electron systems [1], ranging from heavy-fermion [2,3], high-temperature cuprate [4], iron-based pnictide and chalcogenide [5], to the newly-discovered nickelate [6] and graphene-superlattice [7] families. New unusual superconducting phases continue to be discovered, such as those reported during the last two decades in the ferromagnets UGe$_2$, URhGe, and UCoGe [8,9,10]. Instead of antiferromagnetic fluctuations, which are suspected to play a role in most heavy-fermion superconductors [3], ferromagnetic fluctuations were proposed to drive the pairing mechanism of these materials close to a ferromagnetic quantum instability. In these three systems, a magnetic field also leads to a re-entrance or reinforcement of superconductivity, and magnetic-field-induced ferromagnetic fluctuations are suspected to directly control the pairing strength, which can be qualitatively understood as the enhancement of a 'strong-coupling' superconducting parameter $\lambda$ with field [11].

In UGe$_2$ under pressure, a magnetic field along the easy magnetization axis **a** leads to a metamagnetic transition between two ferromagnetic phases, in the vicinity of which superconductivity is reinforced, as indicated by a S-shape in the temperature dependence of the superconducting critical field $H_{c2}$ [12]. Reentrance or reinforcement of superconductivity occurs in the isostructural orthorhombic ferromagnets URhGe and UCoGe under a magnetic field applied along their hard magnetic axis **b** [13,14]. In URhGe, field-induced superconductivity coincides with a metamagnetic transition at $\mu_0 H_m = 12$ T, where enhanced magnetic fluctuations [15,16] accompany a sudden rotation of the magnetic moments (from the initial easy direction **c** to the direction **b**) [13]. The Curie temperature vanishes at $H_m$ in a magnetic field **H** ∥ **b**, and a 'wing structure' of the ferromagnetic phase boundary, linked with a quantum critical end point, can be observed in a magnetic field tilted away from **b** [13]. In this system, as in other heavy-fermion materials, a Fermi surface instability is observed at $H_m$, beyond which a polarized paramagnetic (PPM) regime is established [17,18,19]. In UCoGe, a magnetic field along **b** leads to a reinforcement of superconductivity, which is also associated with the suppression of the Curie temperature at $\simeq 15$ T [14]. However, a metamagnetic transition occurs at a much higher field $\mu_0 H_m \simeq 50$ T, in relation with the temperature $T_\chi^{max}$ at which the magnetic susceptibility presents a broad maximum [20]. The strong exchange field in these ferromagnets indicates that a spin-triplet superconducting order parameter with equal-spin pairing may be realized [11], in contrast to most unconventional superconductors, where antiferromagnetic fluctuations are suspected to be the 'glue' for superconductivity and lead to a singlet order parameter. NMR experiments brought microscopic support for such triplet state and they further highlighted the role of magnetic fluctuations [21,22].

Recently, superconductivity was found to develop in the paramagnetic heavy-fermion material UTe$_2$ at temperatures below $T_{sc} = 1.6$ K [23,24]. This system crystallizes in an orthorhombic crystal structure with space group *Immm* (#71, $D_{2h}^{25}$) and it is characterized by an anisotropic magnetic susceptibility [see Figure 1(a)]. For a magnetic field applied along the easy magnetic axis **a**, a large low-temperature magnetic susceptibility and a scaling plot of magnetization data were interpreted as the indication for a nearby ferromagnetic instability [23]. However, no sign of ferromagnetic order has been found down to the lowest temperatures (25 mK) [25]. While magnetic fluctuations were observed by NMR [26] and muon-spin relaxation measurements [25], the ferromagnetic and/or antiferromagnetic nature of these fluctuations was not determined so far. By analogy with the above-mentioned ferromagnetic superconductors, a spin-triplet nature of superconductivity has been proposed for UTe$_2$ [23]. This proposition was made following the observation of i) a large anisotropic upper critical field which exceeds the normal paramagnetic limitation for all field directions [23,24], ii) a tiny change in the NMR Knight shift through $T_{sc}$ [27], and iii) chiral edge states possibly identified in the superconducting gap by STM experiments [28]. A magnetic field applied along the hard-magnetic axis **b** induces a first-order metamagnetic transition at $\mu_0 H_m \approx 35$ T, which separates a low-field correlated paramagnetic (CPM) regime from a polarized paramagnetic regime [29,30,31]. It is accompanied by sudden jumps $\Delta M \approx 0.3\text{-}0.6\ \mu_B/U$ in the magnetization [29,31] and $\Delta\rho \approx 100\ \mu\Omega$ cm in the residual resistivity [30], and by a large enhancement of the effective mass [29,30,32]. The empirical and almost universal relation 1 T ↔ 1 K between $H_m$ and the temperature $T_\chi^{max} \approx 35$ K at the maximum in the magnetic susceptibility [33], also observed for a large number of heavy-fermion paramagnets [34], indicates that the CPM regime delimited by $H_m$ and $T_\chi^{max}$ is, within a first approximation, controlled by a single energy scale. For **H** ∥ **b**, superconductivity is reinforced above 15 T and it abruptly disappears in the polarized paramagnetic regime above $H_m$ [31,35]. In the following, we will label SC1 and SC2 the respective low-field and high-field regions of the superconducting phase for **H** ∥ **b**. Calorimetric studies showed the appearance under pressure of a second superconducting phase in zero magnetic field, whose critical temperature $T_{sc}$ reaches 3 K at a pressure $p \approx 1.2$ GPa [36]. The extrapolation of the boundary between these two superconducting phases from TDO measurements under pressure and magnetic field [37] may indicate a link between the superconducting phase induced under pressure and the ambient pressure superconducting region SC2. However, to date there is no definitive experimental evidence of a transition between two different superconducting phases SC1 and SC2 at ambient pressure. Alternatively, the upturn in the critical field, which is controlled by a tight balance between the orbital limitation of $H_{c2}$ and the increase of the pairing strength with field [35], could also be induced by a smooth increase of the strong coupling constant $\lambda$.

Figure 1(b) presents a combination of low-temperature magnetic-field versus field-angle phase diagrams of UTe$_2$ obtained in Refs. [31,35]. It summarizes the effect of magnetic fields applied in the (**a**,**b**) and (**b**,**c**) planes. A key property is that the metamagnetic field $H_m$ has a minimal value for **H** ∥ **b**. It strongly increases when the field is tilted from **b** towards the easy magnetic axis **a**, and exceeds the maximum measured field (60 T) for $\phi = $ (**b**,**a**) > 20 °. The increase of $H_m$ is softer when the field is tilted from **b** towards **c**, where it can be followed up to angles $\theta = $ (**b**,**c**) $\approx$ 50 °. At small angles $\phi$ and $\theta$, the field-reinforcement of superconductivity rapidly disappears, and the superconducting critical field shows an almost step-like decrease from 35 T to $\simeq$ 15 T. For larger angles $H_{c2}$ decreases smoothly reaching values of 6 and 10 T for **H** ∥ **a** and **H** ∥ **c**, respectively. A similar suppression of the field-reinforced superconducting phase was reported by tilting the magnetic field away from **b** in UCoGe [11]. For the three field-directions **a**, **b**, and **c**, the low-temperature critical fields $\mu_0 H_{c2,a} \approx 6$ T, $\mu_0 H_{c2,c} \approx 10$ T, and $\mu_0 H_{c2,b} \approx 15\text{-}20$ T (i.e., the extrapolated value of $\mu_0 H_{c2,b}$ ignoring the field-reinforcement below 300 mK) delimiting the low-field superconducting phase SC1 are inversely-correlated with the low-temperature magnetic susceptibilities $\chi_a > \chi_c > \chi_b$ (see Figure 1 and [23,29,33]). A similar inverse relation between the magnetic anisotropy and the anisotropy of $H_{c2}$ was observed in other heavy-fermion superconductors, as URu$_2$Si$_2$ [38,39], CeCoIn$_5$ [40,41], UCoGe and URhGe [11,42]. Spectacularly, a second field-induced superconducting phase was reported in UTe$_2$ for a field direction tilted from **b** towards **c** by an angle $\theta$ ranging from 20 ° to 40 ° [31]. This phase, labelled here as SC-PPM, was observed only in the PPM regime, in fields higher than $\mu_0 H_m \simeq 40\text{-}45$ T, and up to a critical field of $\simeq$ 60 T [31].

In the present work, we focus on a systematic study of the superconducting phases induced in UTe$_2$ at ambient pressure, under a magnetic field applied either along **b**, or tilted by an angle $\theta \simeq 27 \pm 5$ ° from **b** towards **c**. We benefited from a unique combination of extreme conditions offered at the LNCMI-Toulouse, allowing high-magnetic-field electrical resistivity measurements under almost isothermal conditions: long-duration (rise = 70 ms, fall = 300 ms) pulsed magnetic fields up to 60 T, combined with temperatures down to 200 mK. Our results in UTe$_2$ unambiguously show zero-resistance in the SC-PPM phase, confirming its superconducting nature. We extracted the full magnetic-field-temperature phase diagrams of UTe$_2$ for **H** ∥ **b** and **H** tilted by $\theta \simeq 27$ ° from **b** to **c**. From a Fermi-liquid analysis of the resistivity we determine the field dependence of the residual resistivity $\rho_0$ and estimate the variation of the effective mass $m^*$ [43]. These quantities show striking similarities for the two field-directions in contrast with the very different superconducting phase diagrams. In the discussion, we give some elements for the theoretical challenge to understand the nature of these two field-induced superconducting phases in UTe$_2$.

**Results**

*Low-temperature and high-magnetic-field electrical resistivity*

The magnetic-field variation of the electrical resistivity $\rho$ of UTe$_2$ single crystals, measured with a current injected along the **a**-direction, is presented in Figure 2. Data obtained for the two magnetic field directions, **H** ∥ **b** and **H** tilted by $\theta = 27 \pm 5$ ° from **b** in the (**b**,**c**) plane are shown in Figure 2(a-b) and Figure 2(c-d), respectively, for a large range of temperatures varying from 200 mK to 80 K. A comparison of field-up and field-down data (see Supplemental Information) shows almost no heating of the samples by eddy currents in our low-temperature data, which were obtained in long-duration pulsed magnetic fields. At temperatures from $T$ = 2.2 K to $T_{CEP} \approx$ 5-6 K, at which a critical end-point is observed in the present set of data, and under magnetic fields **H** ∥ **b** and **H** tilted by $\theta = 27 \pm 5$ °, similar and sharp first-order step-like increases of $\rho$ are observed at the metamagnetic field $\mu_0 H_m$, which equals 34 and 45 T for the two field directions, respectively. For both directions, when the temperature is increased above $T_{CEP}$, the sharp anomaly at $H_m$ is transformed into a broad maximum, at a field also labeled $H_m$, which vanishes at temperatures higher than 30 K. Below we focus on the signatures of superconductivity in the low-temperature data.

Figure 2(b) shows that, for **H** ∥ **b**, field-induced superconductivity develops just below $H_m$, with an onset at a maximal temperature of 1.2 K and a zero-resistivity reached below the maximal superconducting temperature $T_{SC} \simeq 1$ K. In spite of a non-zero resistivity due to small out-of-phase contamination of the signal, this new set of data confirms, in magnetic fields extended up to 60 T, the two recent reports of field-reinforcement of superconductivity in UTe$_2$ for **H** ∥ **b** [31,35]. For **H** tilted by $\theta = 27 \pm 5$ ° from **b** in the (**b**,**c**) plane, Figure 2(d) shows unambiguously a zero-resistivity regime in fields higher than $H_m$, whereas zero resistance was not obtained in the pioneering work [31], possibly due to heating and/or out-of-phase contamination of the signal. These data support the presence of a field-induced superconducting phase SC-PPM above $H_m$ [31]. After an onset at a maximal temperature of 2 K, zero-resistivity is reached below the maximal superconducting temperature $T_{SC} \simeq 1.5$ K, which is higher than the superconducting temperature reported at any field applied along **b**. The magnetic field at which the zero-resistivity superconducting phase SC-PPM develops is locked to the value $\mu_0 H_m \simeq 45$ T observed for $T > T_{SC}$. Inside the CPM regime, the onset of the phase SC-PPM at $\simeq 43$ T precedes the zero-resistivity-state reached beyond $H_m$. We also confirm that the low-field superconducting phase SC1 is well-separated from the field-induced phase SC-PPM. At the lowest temperature, the phase SC1 vanishes at a moderate critical field of $\simeq 10$ T.

*Temperature-magnetic field phase diagrams and quantum critical fluctuations*

Figure 3(a) presents the magnetic-field-temperature phase diagram extracted here for UTe$_2$ in a field **H** || **b**. The phase diagram shows two domes corresponding to the superconducting regions SC1 at low-field and SC2 induced by a magnetic field. The transition temperature $T_{SC}$ of SC2 is maximal at a magnetic field just below $\mu_0 H_m$ = 34 T. SC2 is presumably driven by the magnetic fluctuations induced on approaching the metamagnetic transition, which also control the enhancement of the Sommerfeld coefficient $\gamma$ in the heat capacity [34] and of the coefficient $A$ of the Fermi liquid $T^2$ term of the electrical resistivity [30]. We confirm here that SC2 is strictly bounded by $H_m$, at which the magnetization was found to jump from 0.4 to 1 $\mu_B$/U and above which a PPM regime is reached [29].

Figure 3(b) presents the magnetic field - temperature phase diagram extracted here for UTe$_2$ in a field **H** tilted by $\theta$ = 27 ± 5 ° from **b** in the (**b**,**c**) plane. While the low-field superconducting phase SC1 vanishes at a critical field $H_{c2} \simeq$ 10 T, $\mu_0 H_m$ reaches 45 T at low temperature for this field direction. When the temperature is increased, the behavior is similar to that reported for **H** || **b**: $H_m$ loses its first-order character at the temperature $T_{CEP} \approx$ 5-6 K. It transforms into a cross-over at higher temperatures and finally disappears above 20-30 K. In agreement with the previously-published data [31], the superconducting phase SC-PPM is only observed in fields higher than $H_m$, and up to a superconducting critical field higher than 60 T at low temperature. A maximal field-induced superconducting temperature $T_{SC} \approx$ 1.5 K appears at a field close to $H_m$, emphasizing a direct link with the metamagnetic transition.

In many heavy-fermion magnets, a maximum of the effective mass is observed in the vicinity of a magnetic instability. It is commonly understood as resulting from the quantum critical magnetic fluctuations, coupled or not with a Fermi surface instability [44]. Within a Fermi-liquid description, the electrical resistivity can be fitted by $\rho(T) = \rho_0 + AT^2$, and the $A$ coefficient varies as the square of the effective mass $m^*$. In heavy-fermion systems, $m^*$ is mainly controlled by magnetic fluctuations related with the proximity of quantum magnetic instabilities. We note that considering the coefficient $A$ can lead to an overestimation of $m^*$ [45]. Figures 3(a) and 3(b) present the magnetic-field variations of $A$ and $\rho_0$, respectively, extracted here for UTe$_2$ with **H** || **b** and **H** tilted by $\theta$ = 27 ± 5 ° from **b** to **c**. Fermi-liquid-like fits to the high-field resistivity data were done for all fields investigated here, in the temperature windows 1.5 ≤ $T$ ≤ 4.2 K for **H** || **b**, and 2.2 ≤ $T$ ≤ 4.2 K for **H** tilted by $\theta$ = 27 ° (see Supplemental Material). We find almost similar field-variations of $A$ and $\rho_0$ for the two field directions: at $H_m$, while $A$ increases by a factor $\simeq$ 6 and passes through a sharp maximum, $\rho_0$ undergoes a sharp step-like enhancement, jumping from 15 to 80 $\mu\Omega$.cm. The field-variation of $A$ reported here for **H** || **b** is in good agreement with a previous report [30], and it indicates a sharp and strong enhancement of the magnetic fluctuations at $H_m$. For **H** || **b**, a qualitatively similar enhancement of $m^*$ at $H_m$ was found by applying a Maxwell relation to magnetization data [29] and by direct heat-capacity measurements [32].

Differences between the two field-directions are visible from plots of $A$ and $\rho_0$ versus $H/H_m$ [Figures 4(a) and 4(b)]. While the variation of $A$ through $H_m$ is almost symmetric for **H** || **b**, it is slightly asymmetric for **H** tilted by $\theta$ = 27 ± 5 ° from **b**. For the tilted-field direction, $A(H)$ is steeper for $H < H_m$ and more gradual for $H > H_m$. As well, the decrease of $\rho_0$ beyond $H_m$ is more marked for **H** tilted by $\theta = 27 \pm 5$ ° from **b**. New high-field experiments are now needed for a complete angular study of the Fermi-liquid behavior.

**Discussion**

The ultimate goal would be to provide a full microscopic description of the different superconducting phases and their pairing mechanisms in UTe$_2$. We are still far from this objective, but the experimental data presented here, in complement to those from [31], offer a broad set of constraints for theories. A striking feature of the phase diagrams presented in Figures 3(a-b) is that the superconducting phases SC2 for **H** || **b** and SC-PPM in a field **H** tilted by $\theta = 27 \pm 5$ ° from **b** towards **c** are bounded by the metamagnetic field $H_m$, with a substantial difference that the phase SC2 is pinned inside the CPM regime and it does not survive in the PPM regime while, inversely, the phase SC-PPM is pinned inside the PPM regime and does not develop in the CPM regime. A natural explanation would be that the pairing mechanism changes drastically on crossing the first-order line $H_m$, at which one would expect a difference in the nature of the critical magnetic fluctuations in the CPM and PPM regimes. This difference would change substantially for the two field-directions **H** || **b** and **H** tilted by 27° from **b**.

A rough estimation of the field-dependence of the pairing strength can be obtained from the Fermi-liquid analysis done above, where a maximum of the quadratic coefficient $A$ at the metamagnetic transition indicated an increase of the effective mass $m^*$, presumably controlled by critical magnetic fluctuations. In a simple picture, the effective mass can be related by $m^* \sim 1+\lambda$ to the superconducting pairing strength $\lambda$ [46,47]. However, the fact that the enhancement of $A$ is almost symmetric around $H_m$ is puzzling with respect to the abrupt suppression of superconductivity for **H** || **b** , and its abrupt appearance for **H** tilted by $\theta = 27 \pm 5$ ° from **b** towards **c**. A similar symmetrical enhancement of $A$ has been observed at the metamagnetic transition in other heavy fermion systems, where a drastic change of magnetic fluctuations and Fermi surfaces was found [48,49]. The abrupt disappearance/appearance of superconductivity at $H_m$ could also result from a sudden change of the Fermi surface. A Fermi surface reconstruction is compatible with the large and sudden variation of the residual resistivity at $H_m$ for the two field directions, but also with the sign changes in the thermoelectric power and Hall coefficient at $H_m$ for **H** || **b** [50]. However our results raise a serious hurdle to both these pictures since the field-driven enhancement of $A$ is very similar for **H** || **b** and **H** tilted by $\theta = 27 \pm 5$ ° from **b** to **c**. The asymmetry in the field-variation of $A$ for **H** tilted by 27 ° suggests that the magnetic fluctuations may be slightly more intense above $H_m$ for this field direction, but this effect is too small to explain the differences between the phases SC2 and SC-PPM. The magnetization jump at $H_m$ is also very similar for **H** || **b** and **H** tilted by 27 ° [31]. Extra ingredients are, thus, needed to describe the field and angle domains of stability of these two field-induced superconducting phases. In the following, we mention elements that may be considered for such description.

Figure 5 presents views of the crystal structure of UTe$_2$ where the magnetic uranium ions can be seen to form a ladder structure [51]. We highlight the family of reticular (and cleaving) planes of Miller indices (0 1 1), which contain sets of ladders having the smallest inter-ladder U-U distance ($d_3 = 4.89$ Å). Interestingly, the direction **n** normal to these planes coincides, within the experimental uncertainty, with the field-direction along which the phase SC-PPM develops [31]. It lies in the (**b**,**c**) plane and has an angle $\theta = 23.7$ ° with **b** (see Supplementary Information). Although the connection with the pairing mechanism remains unclear, this coincidence may not be accidental and may constitute a possible line of approach for future theories.

In relation with the ladder structure, magnetic frustration has been invoked as a possible origin of the paramagnetic ground state in $UTe_2$ at zero field and ambient pressure, and a competition between ferromagnetic and antiferromagnetic configurations has been discussed [51,52]. Electronic structure calculations pointed out that the ground state is sensitive to the Coulomb repulsion, and that the ferromagnetic and antiferromagnetic configurations are energetically-close [51]. The respective roles of ferromagnetic and antiferromagnetic fluctuations in $UTe_2$ may, thus, be important for the superconducting phases, and this question needs to be clarified. While $UTe_2$ was first proposed to be nearly-ferromagnetic [23], the nature of the pressure-induced magnetic phase, initially reported in [36], was not determined so far. Several studies suggested that $UTe_2$ is not a simple ferromagnet and may be close to an antiferromagnetic instability [53,54,55]. At ambient pressure, the absence of metamagnetism in a magnetic field up to 55 T applied along the easy magnetic axis **a** [29,30] indicates that $UTe_2$ is at least not a conventional Ising paramagnet close to a ferromagnetic instability, unlike $UGe_2$ under pressure [56] and $UCoAl$ at ambient pressure [57]. The negative Curie-Weiss temperatures extracted from the high-temperature magnetic susceptibility, for the three directions **H** || **a**, **b**, and **c** [see Figure 1(a)], indicate antiferromagnetic exchange interactions (see also [33,58]). A broad maximum at the temperature $T_\chi^{max} = 35$ K in the magnetic susceptibility for **H** || **b** is also compatible with the onset of antiferromagnetic fluctuations, as observed in several heavy-fermion paramagnets [34]. Low-temperature downward deviations of the magnetic susceptibility for **H** || **a,c** (in comparison with its high-temperature behavior) are observed in the log-log plot shown in Inset of Figure 1(a). These deviations confirm the formation of a heavy-fermion state below 50 K, which may coincide with the onset of antiferromagnetic fluctuations. Interestingly, the high-temperature magnetic susceptibility for **H** || **a** varies as $1/T^{0.75}$ over more than one decade, from 20 to 300 K. However, further investigations would be needed to understanding this power-law behavior.

The different superconducting regimes may correspond to different order parameters, with different sensitivities to a magnetic field. It has been generally assumed that all the superconducting phases in $UTe_2$ have a triplet order parameter, mainly because of high values of the superconducting upper critical field, a small decrease of the NMR Knight shift below $T_{SC}$ [27] and a supposed proximity to ferromagnetism [23,24,51,59]. However, this still needs confirmation especially if, as pointed out above, antiferromagnetic fluctuations may play a much larger role than initially thought. The disappearance of superconducting phase SC2 as the PPM regime is entered for **H** || **b** could indicate that some paramagnetic limiting effect is present. Thereafter, for **H** tilted by $\theta = 27 \pm 5$ ° from **b** to **c** the phase SC-PPM could be a natural candidate for triplet superconductivity with no paramagnetic limitation. However, two questions remain: why this phase appears only for such a specific angular range, possibly in relation with the previous symmetry considerations, and especially why this phase does not develop in fields smaller than $H_m$?

A full understanding of the magnetic fluctuations and their feedback on the superconducting pairing undoubtedly requires the knowledge of the Fermi surface and electronic structure of $UTe_2$. As mentioned above, calculated Fermi surfaces strongly depend on the Coulomb repulsion $U$ [51,52,60]. Two-dimensional Fermi surfaces along **c**, similar to that of $ThTe_2$ and corresponding to a localized $f$-electrons limit, are expected for large values of $U$ [51,52,61]. For quasi one-dimensional [62] or quasi two-dimensional [63] Fermi surfaces, it is predicted that the orbital limit could be suppressed for particular field directions, which may help explaining both superconducting regions SC2 and SC-PPM. However, the validity of such models was not proven so far, and no evidence for low-dimensional features in $UTe_2$ was found from bulk properties. An alternative scenario to describe the absence of paramagnetic limitation could be based on the combination of a Jaccarino-Peter compensation with a field dependent pairing strength in the polarized phase. Such analysis was recently proposed to reproduce the critical superconducting fields of UTe$_2$ in the (**b**,**c**) plane [64].

Although the measurements presented here and in other works start to bring a clear picture of the complex phase diagram of UTe$_2$, which includes multiple superconducting and magnetic phases, we are still far from a deep understanding of its electronic properties. A target is now to perform microscopic studies to identify the nature of the magnetic fluctuations, their change through $H_m$, and the superconducting order parameters. Theoretical developments are also needed to determine the superconducting pairing mechanism(s). This is a stiff challenge but the rare flurry of stunning phenomena observed in UTe$_2$ fully justifies such forthcoming efforts.

**Methods**

**Samples.** Single crystals of UTe$_2$ were prepared by the chemical vapor transport method with similar parameters as described in Ref. [23]. Their structure and orientation was checked by single-crystal X-ray diffraction. A sharp bulk transition at $T_{sc}$ = 1.6 K was indicated from specific heat measurements, while zero-resistivity at temperatures below $T_{sc}$ was confirmed by zero-field AC resistivity measurements.

**Pulsed-field experiments.** Magnetoresistance measurements were performed at the Laboratoire National des Champs Magnétiques Intenses (LNCMI) in Toulouse under long-duration pulsed magnetic fields, either up to 68 T (30 ms raise and 100 ms fall) and combined with an $^4$He cryostat offering temperatures down to 1.4 K, or up to 58 T (55 ms rise and 300 ms fall) and combined by a home-developed dilution fridge made of a non-metallic mixing chamber offering temperatures down to 100 mK. A standard four-probe method with currents **I** ∥ **a**, at a frequency of 20–70 kHz, and a digital lock-in detection were used. Resistivity data were normalized so that the maximal value, at a temperature of ≈ 65 K and at zero-field, reaches 450 μΩ·cm (a different normalization lead to a maximum of 650 μΩ·cm in a previous work [30]). Normalization was made following absolute resistivity measurements on samples whose geometrical shape was known.

**Data availability.** Data and materials concerning the experiments can be available by directly contacting W. Knafo (william.knafo@lncmi.cnrs.fr.).

**Acknowledgments**

We acknowledge A. Miyake, J. Béard, F. Hardy, J.-P. Brison, K. Ishida, Y. Tokunaga, Y. Yanase, and H. Harima for useful discussions.

This work at the LNCMI was supported by the "Programme Investissements d'Avenir" under the project ANR-11-IDEX-0002- 02 (reference ANR-10-LABX-0037-NEXT). We acknowledge the financial support of the Cross-Disciplinary Program on Instrumentation and Detection of CEA, the French Alternative Energies and Atomic Energy Commission, and KAKENHI (JP15H05882, JP15H05884, JP15K21732, JP16H04006, JP15H05745, JP19H00646).

**Author contributions**

Samples were grown by G.L in close collaboration with D.A. They were characterized in zero and low fields by G.L., M.V., D.B. and G.K. Samples measured in pulsed fields were prepared by M.V.

Experiments in pulsed magnetic field were performed by W.K., M.N., and A.Z.. Data were analyzed by W.K. The paper was written by W.K. and D.B., with contributions from all of the authors.

**Competing financial interests:** There are no competing financial interests.

# Figures

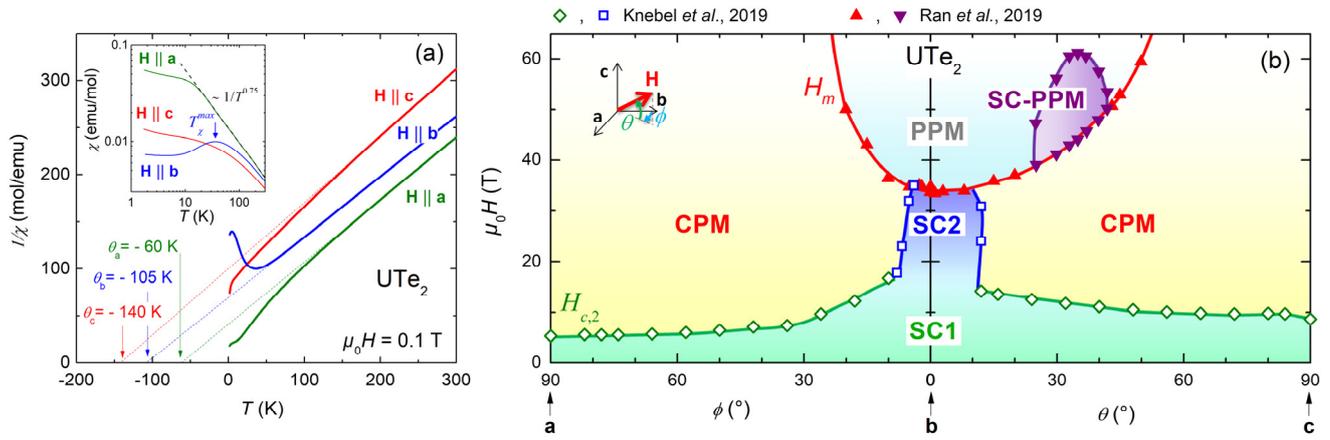

**Figure 1. Magnetic susceptibility and phase diagram of UTe$_2$.** (a) Temperature-dependence of the inversed magnetic susceptibility $1/\chi$ of UTe$_2$ in magnetic fields **H** applied along the three main crystallographic directions **a**, **b**, and **c**. Inset: Temperature-dependence of the magnetic susceptibility $\chi$ for **H** ∥ **a**, **b**, and **c**, in a log-log scale. (b) Low-temperature magnetic phase diagram of UTe$_2$, in fields applied along variable directions from **b** to **a** (angle $\phi$) and from **b** to **c** (angle $\theta$). Data from Refs. [31,34] were plotted in this Figure.

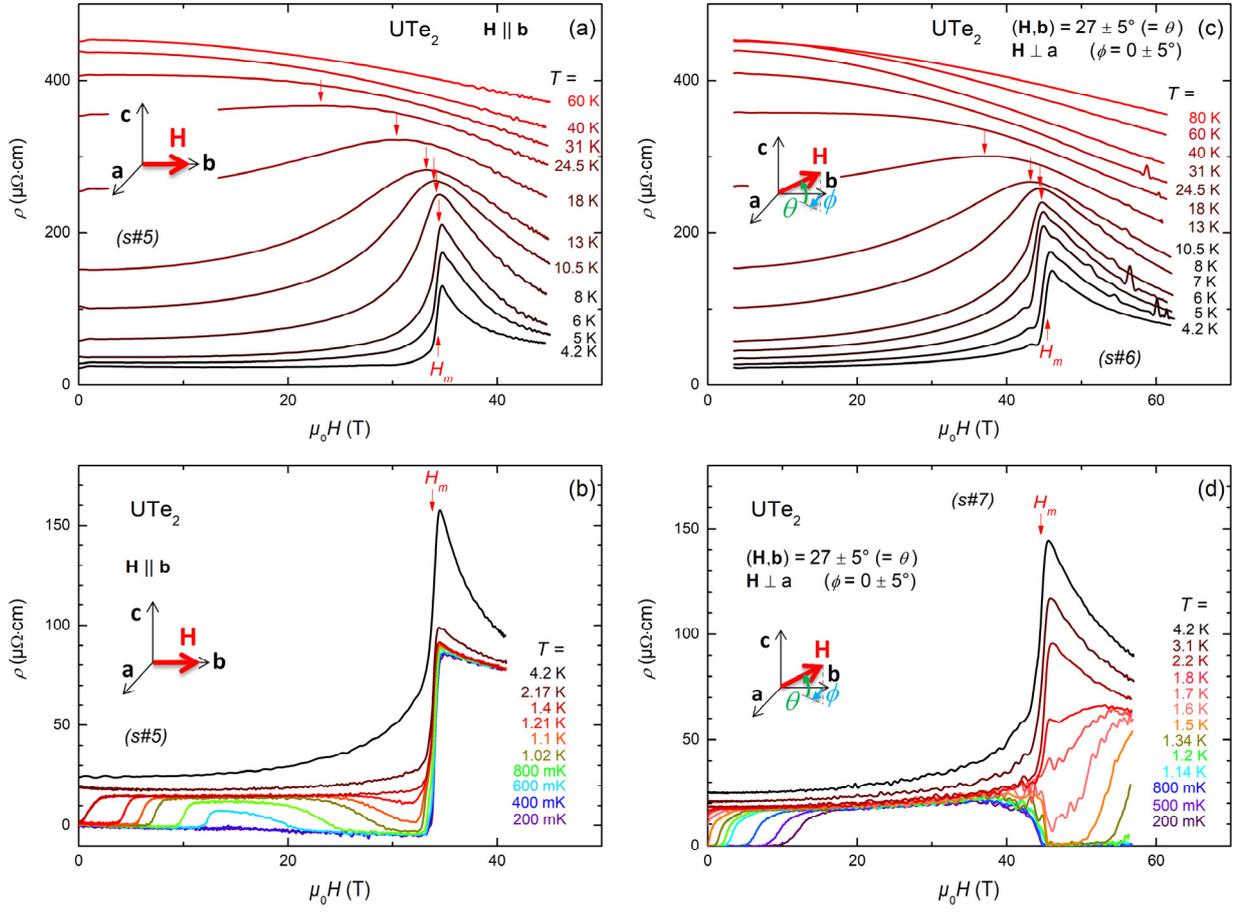

**Figure 2. Electrical resistivity of UTe$_2$ versus magnetic field.** (a) High-temperature and (b) low-temperature resistivity of UTe$_2$ in a magnetic field **H**||**b**. (c) High-temperature and (d) low-temperature resistivity of UTe$_2$ in a magnetic field **H** tilted by 27±5 ° from **b** in the (**b**,**c**) plane.

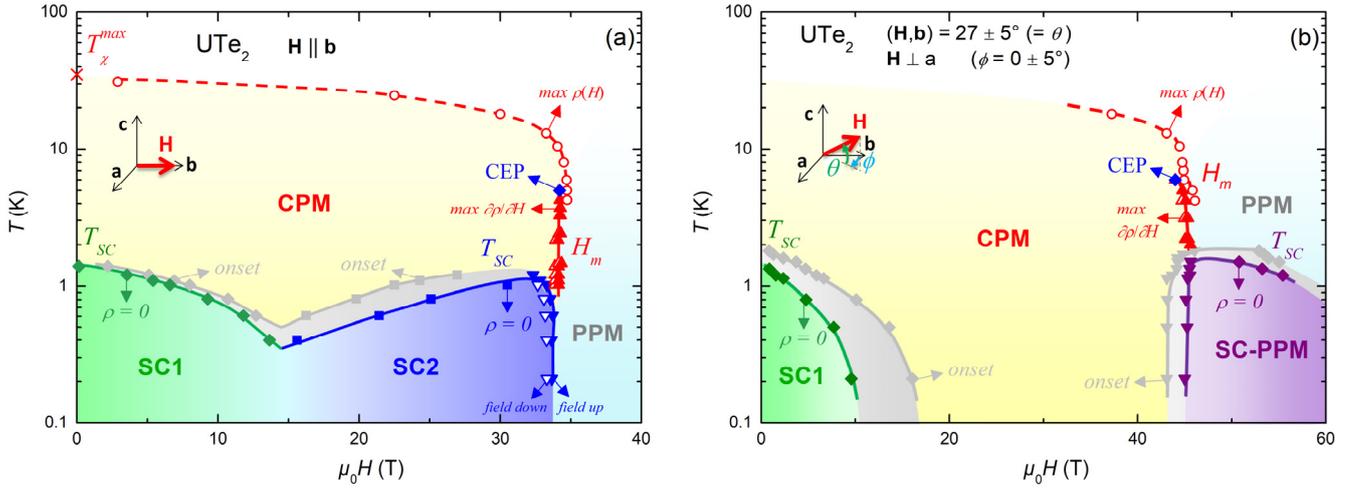

**Figure 3. Magnetic phase diagrams of UTe$_2$.** (a) Magnetic-field-temperature phase diagram of UTe$_2$ in a magnetic field **H**∥**b**. (b) Magnetic-field-temperature phase diagram of UTe$_2$ in a magnetic field **H** tilted by 27±5 ° from **b** in the (**b**,**c**) plane. CPM = correlated paramagnetism, PPP = polarized paramagnetism, SC1, SC2, SC-PPM indicate the different superconducting phases (or regions). CEP indicates the critical end-point of the first-order metamagnetic transition.

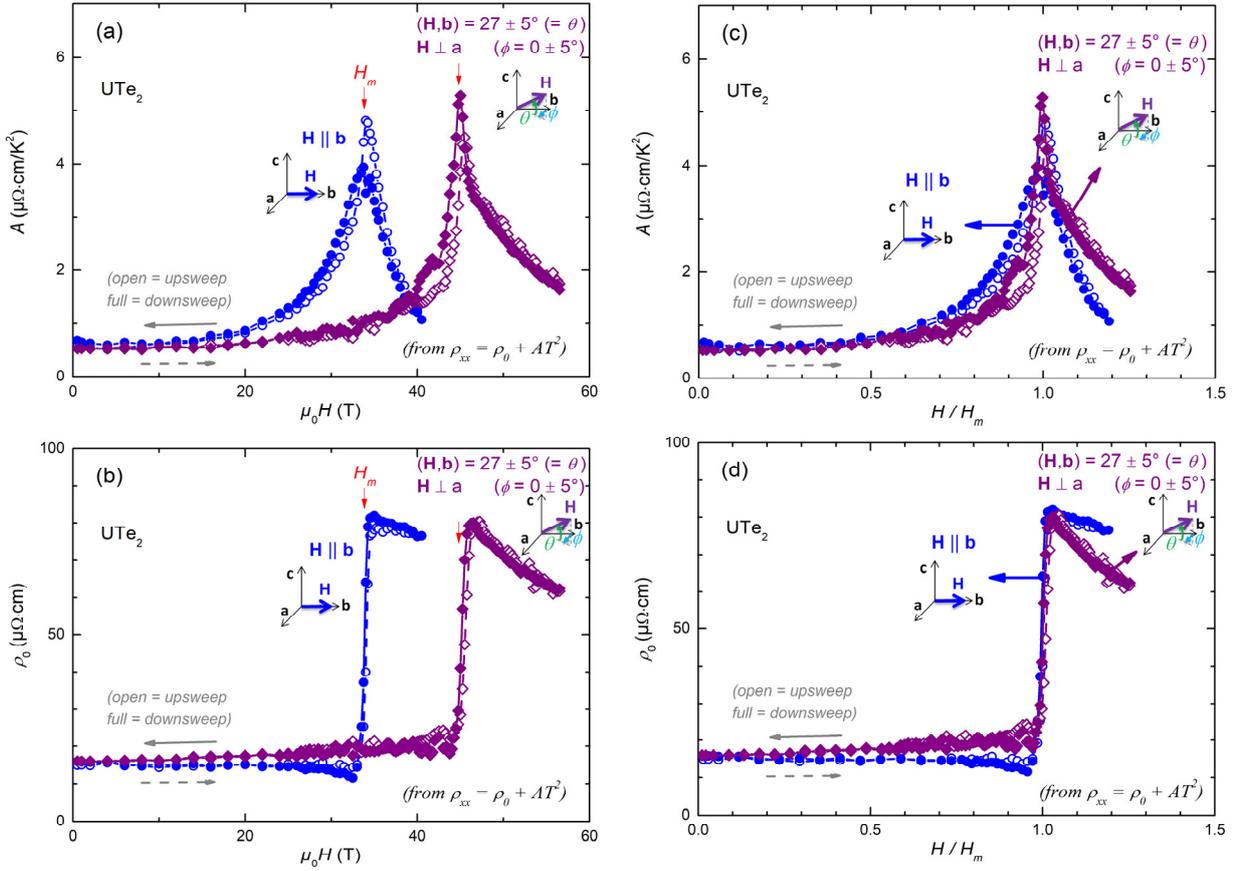

**Figure 4. Quadratic coefficient *A* and residual resistivity of UTe$_2$.** (a) Magnetic-field variation of the quadratic coefficient *A* and (b) residual resistivity $\rho_0$ extracted from Fermi-liquid fits to the electrical resistivity of UTe$_2$ in a magnetic field **H**||**b** and in a magnetic field **H** tilted by 27±5 ° from **b** in the (**b**,**c**) plane. Plots of (c) *A* and (d) $\rho_0$ versus $H/H_m$ for the two field-directions. Data are presented for both field-up and field-down sweeps. Details about the Fermi-liquid fits are shown in Supplementary Figure 7 of the Supplementary Information.

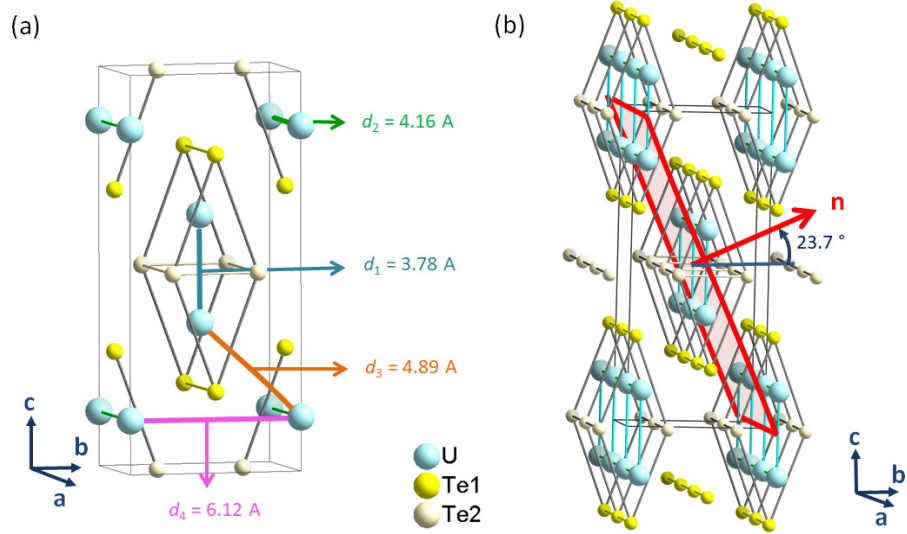

**Figure 5. Crystal structure of UTe$_2$.** (a) Elementary unit cell and identification of 4 exchange paths corresponding to the smallest U-U distances, (b) Extended crystal structure emphasizing the network of two-leg ladders. The vector **n** normal to a family of reticular (and cleaving) planes of Miller indices (0 1 1), with an angle $\theta = (\mathbf{b},\mathbf{n}) = 23.7\,°$, is indicated. These reticular planes are characteristic of the ladder structure.

# Supplementary Information

# Comparison of two superconducting phases induced by a magnetic field in UTe$_2$

W. Knafo,[1,*] M. Nardone,[1] M. Valiska,[2] A. Zitouni,[1] G. Lapertot,[2] D. Aoki,[2,3] G. Knebel,[2] D. Braithwaite[2]

[1] *Laboratoire National des Champs Magnétiques Intenses, UPR 3228, CNRS-UPS-INSA-UGA, 143 Avenue de Rangueil, 31400 Toulouse, France*
[2] *Univ. Grenoble Alpes and CEA, IRIG-PHELIQS, F-38000 Grenoble, France*
[3] *Institute for Materials Research, Tohoku University, Ibaraki 311-1313, Japan*

\* *Corresponding author: william.knafo@lncmi.cnrs.fr*

In Supplementary Figures 1-4, we present complementary plots of the low-temperature resistivity data. In particular, a comparison of field-up and field-down data is shown, indicating almost negligible eddy current heating of the sample in the magnetic field pulses.

Supplementary Figure 5 (a) shows a zoom on the low-temperature resistivity of $UTe_2$ in a magnetic field $\mathbf{H}\|\mathbf{b}$ close to $H_m$. A small negative value of $\rho$ is due to out-of-phase contamination in the resistive signal in high fields. At $T = 1$ and 1.4 K, a hysteresis of field width $\Delta H = 0.25$ T is visible at the first-order transition field $H_m$, which reaches 33.9 and 34.15 T (minimum of slope of $\rho$) for falling and rising fields, respectively. Supplementary Figure 5(b) shows that the hysteresis observed at $H_m$ at $T = 1.8$ and 2.2 K is lost at low temperatures once superconductivity develops. No out-of-phase contamination is observed in this set of data.

Supplementary Figure 6 presents resistivity versus temperature plots in magnetic fields $\mathbf{H} \| \mathbf{b}$ and $\mathbf{H}$ tilted by 27±5 ° from b in the (b,c) plane. In Supplementary Figure 7, the Fermi-liquid like fits to the electrical resistivity data are presented. These fits were used to extract the magnetic-field variations of the quadratic coefficient $A$ and of the residual resistivity $\rho_0$.

Supplementary Figure 8 presents views of the crystal structure and of the Brillouin zone, where the direction $\mathbf{n}$ in the real space, which is equivalent to the direction $\mathbf{k} = (0\ 1\ 1)$ in the reciprocal space, are identified as a peculiar direction close to the direction along which a magnetic field induced the superconducting phase SC-PPM.

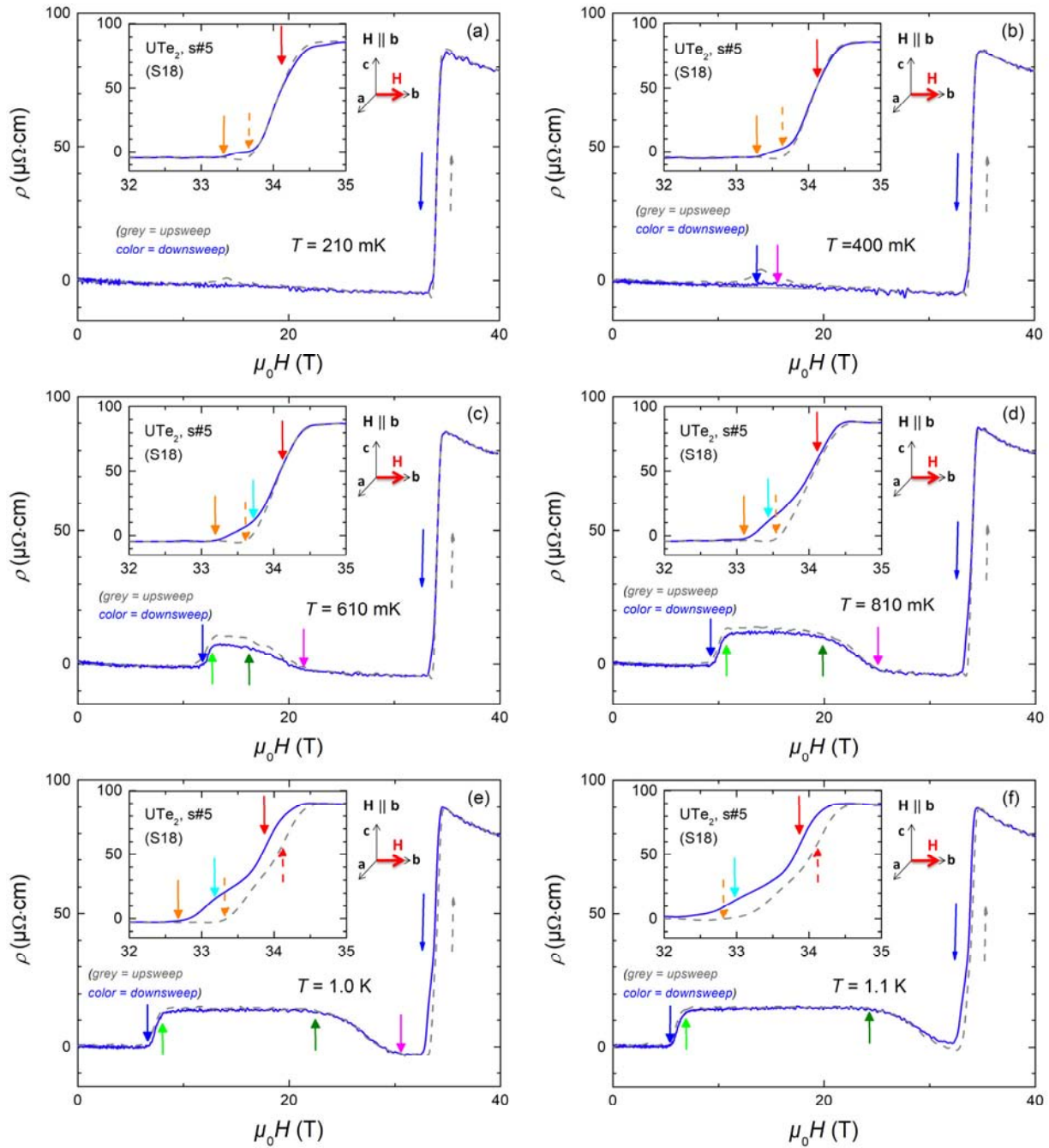

**Supplementary Figure 1. Electrical resistivity of UTe$_2$ in a magnetic field H∥b.** Comparison of field-up and field-down sweeps at temperatures from 210 mK to 1.1 K.

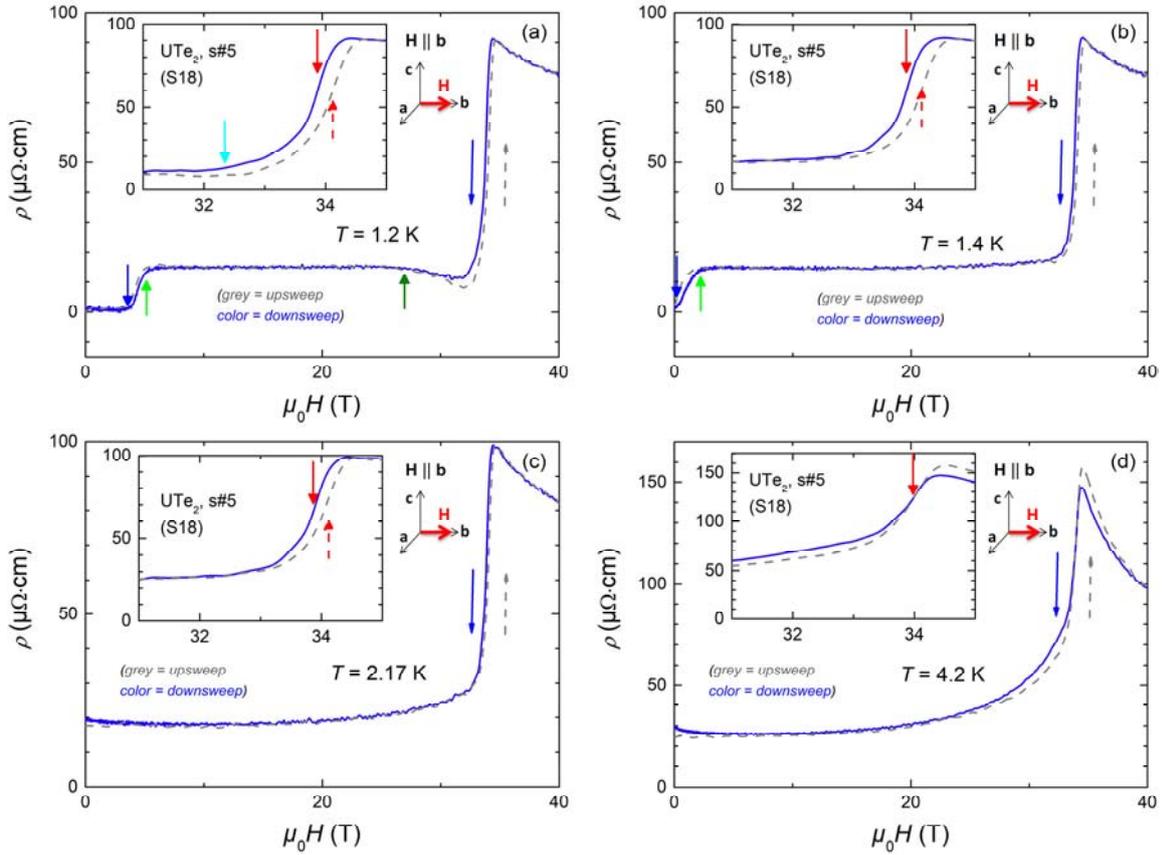

**Supplementary Figure 2. Electrical resistivity of UTe$_2$ in a magnetic field H∥b**. Comparison of field-up and field-down sweeps at temperatures from 1.2 K to 4.2 K.

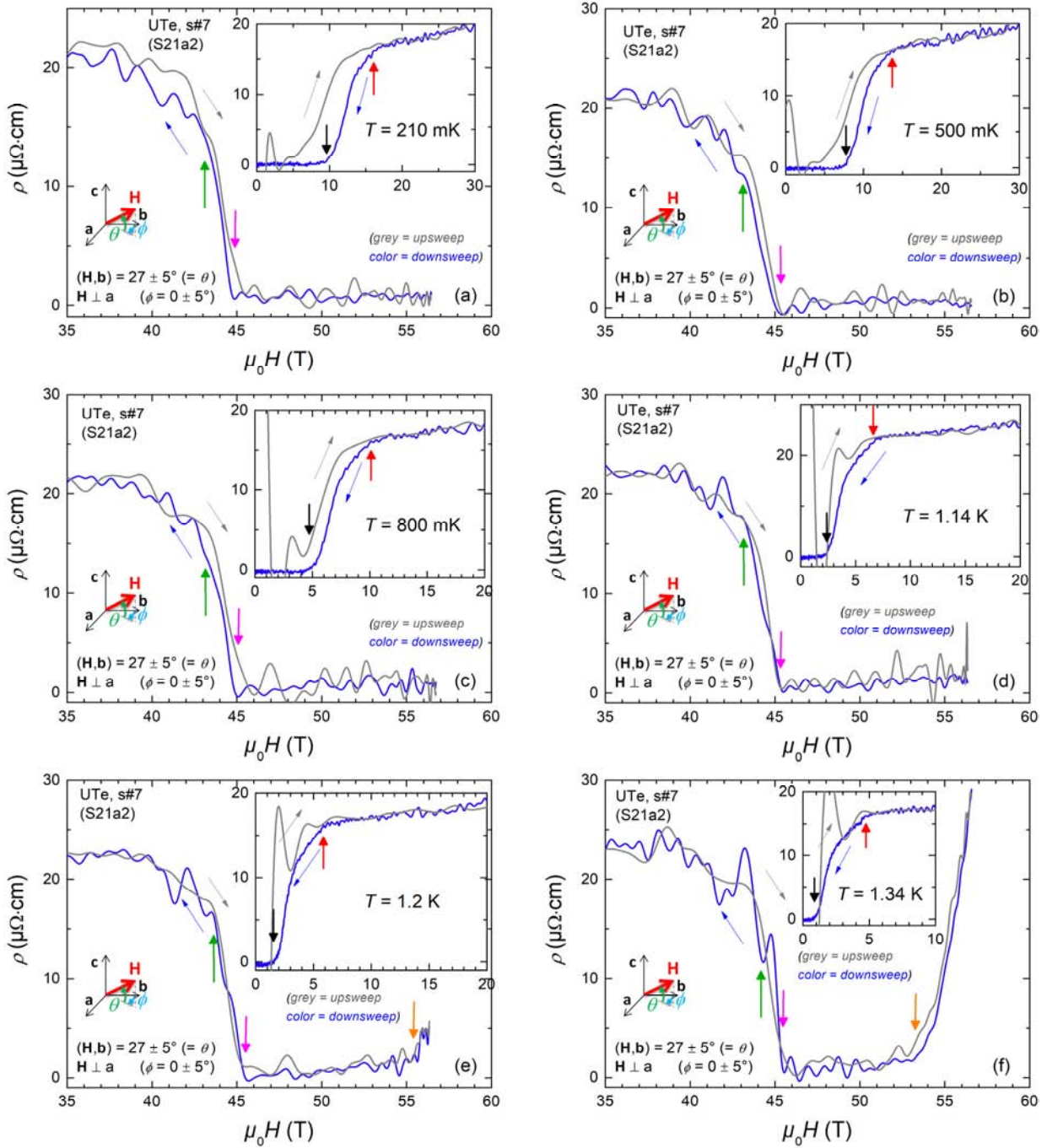

**Supplementary Figure 3. Electrical resistivity of UTe$_2$ in a magnetic field H tilted by 27±5 ° from b in the (b,c) plane**. Comparison of field-up and field-down sweeps at temperatures from 210 mK to 1.34 K.

.

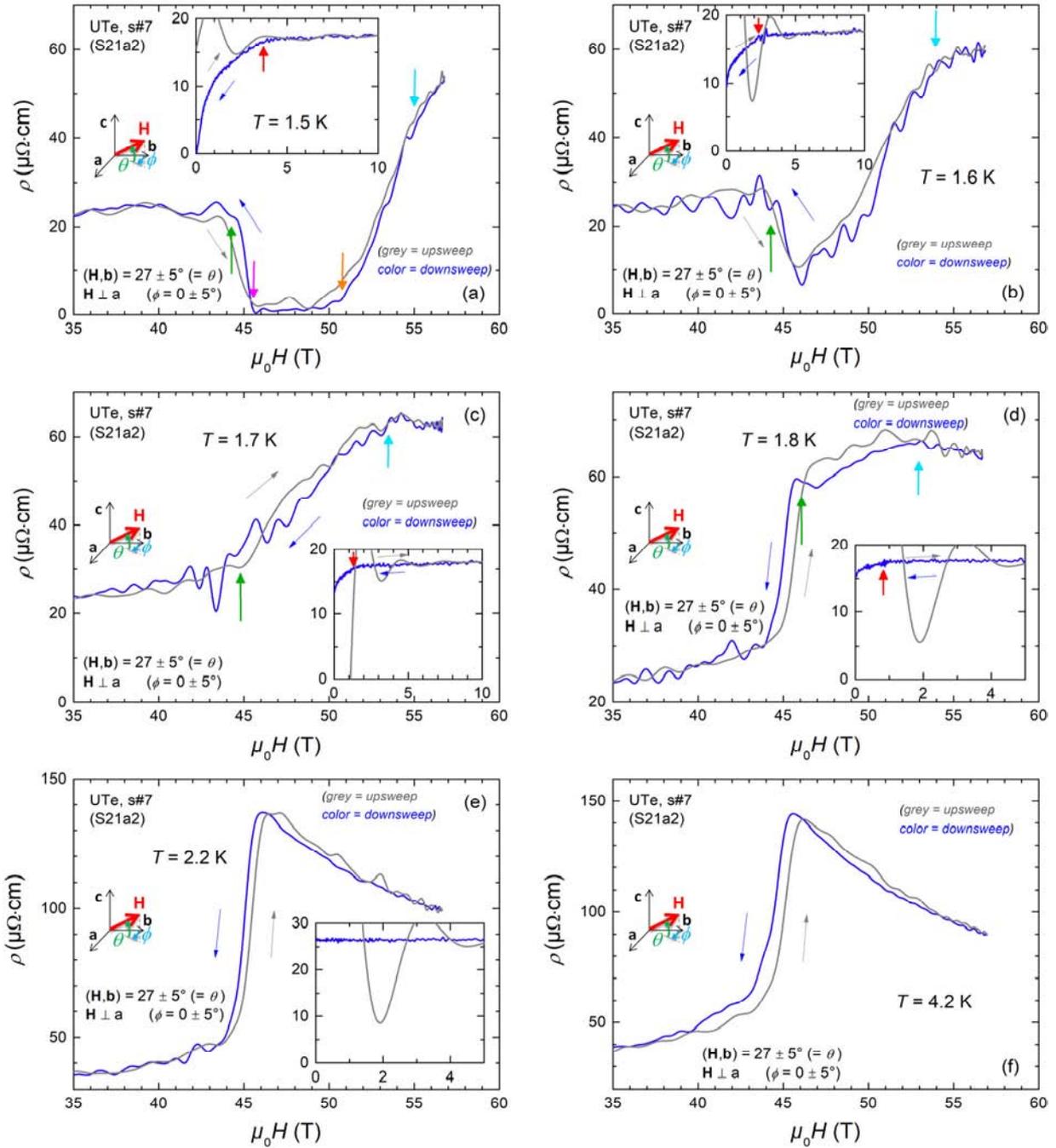

**Supplementary Figure 4. Electrical resistivity of UTe$_2$ in a magnetic field H tilted by 27±5 ° from b in the (b,c) plane.** Comparison of field-up and field-down sweeps at temperatures from 1.5 K to 4.2 K.

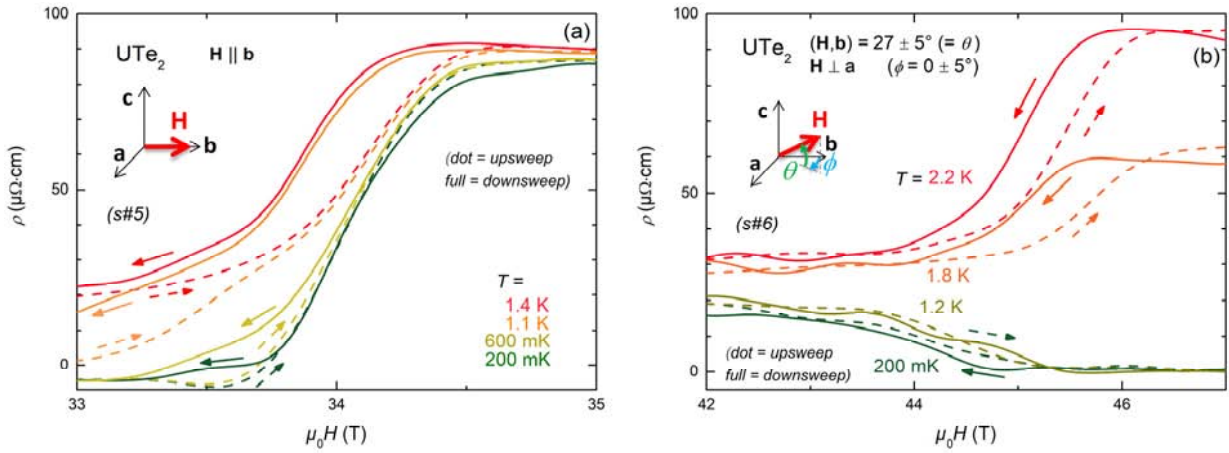

**Supplementary Figure 5. Electrical resistivity of UTe$_2$ in the vicinity of the metamagnetic transition.** (a) Low-temperature resistivity of UTe$_2$ in a magnetic field **H**∥**b**. (b) Low-temperature resistivity of UTe$_2$ in a magnetic field **H** tilted by 27±5 ° from **b** in the (**b**,**c**) plane.

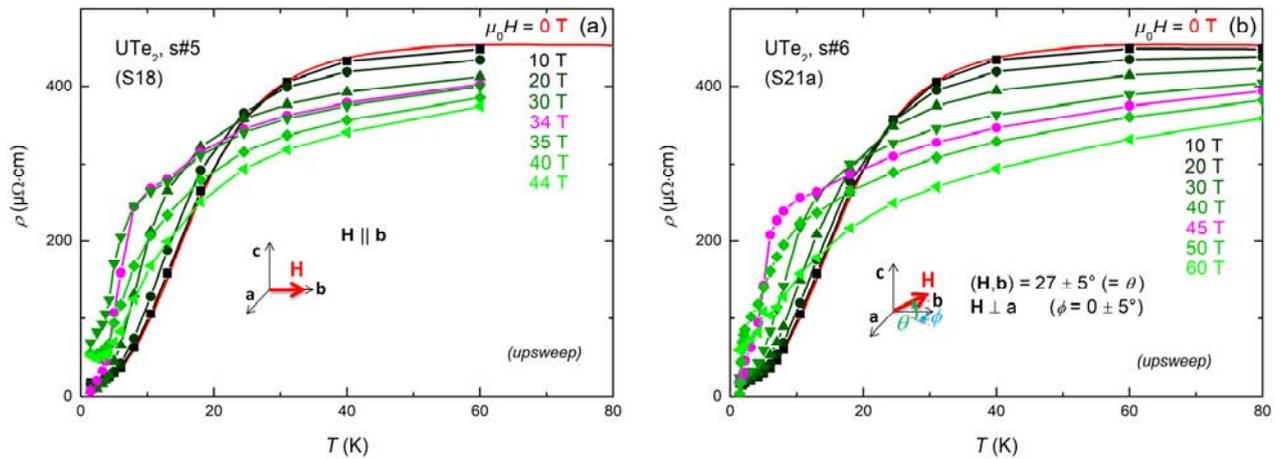

**Supplementary Figure 6. Electrical resistivity of UTe$_2$ versus temperature in magnetic fields H ∥ b and H tilted by 27±5 ° from b in the (b,c) plane.** Data are presented for field-up sweeps.

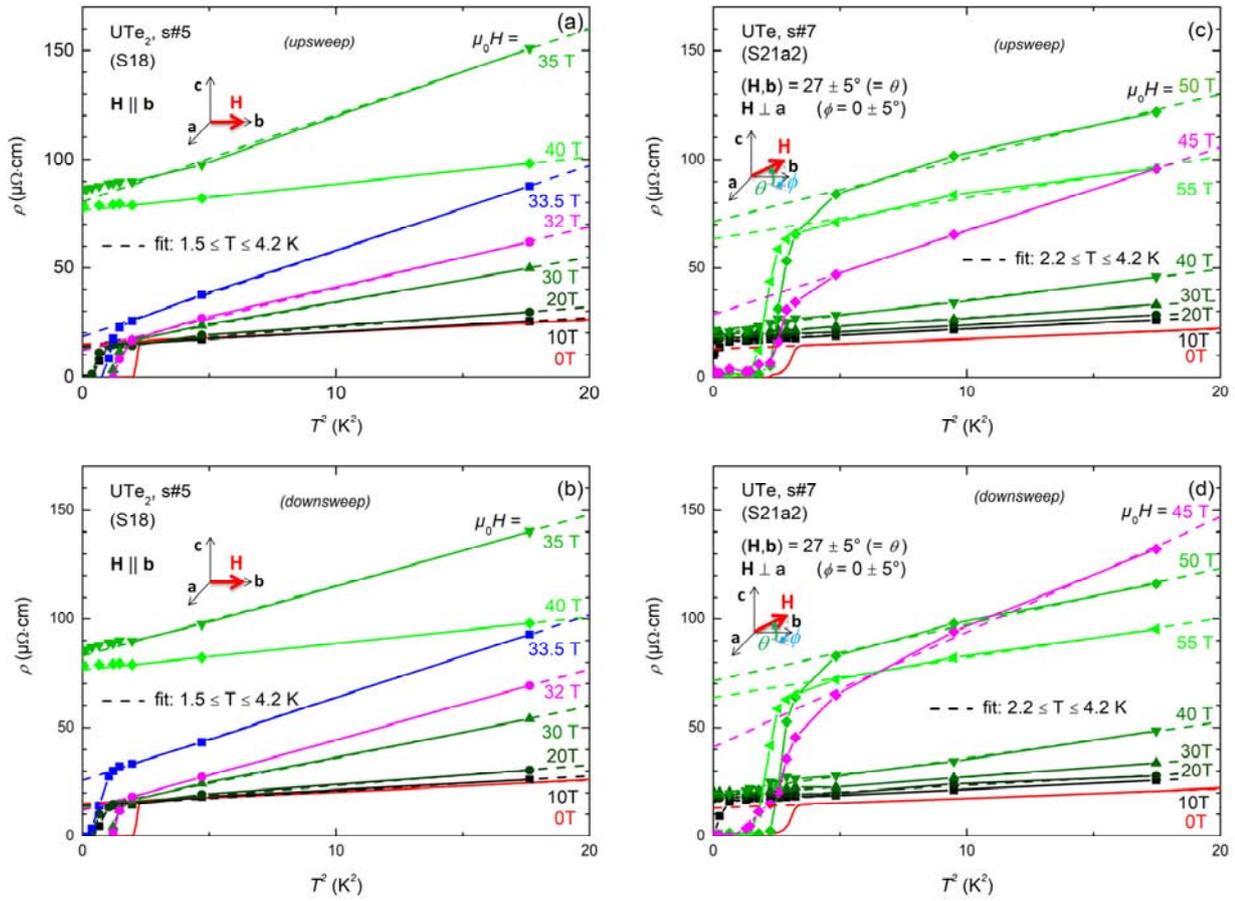

**Supplementary Figure 7. Electrical resistivity of UTe$_2$ versus square of temperature and its T$^2$ fits in magnetic fields H ||b and H tilted by 27±5 ° from b in the (b,c) plane.** Data are presented for field-up and field-down sweeps.

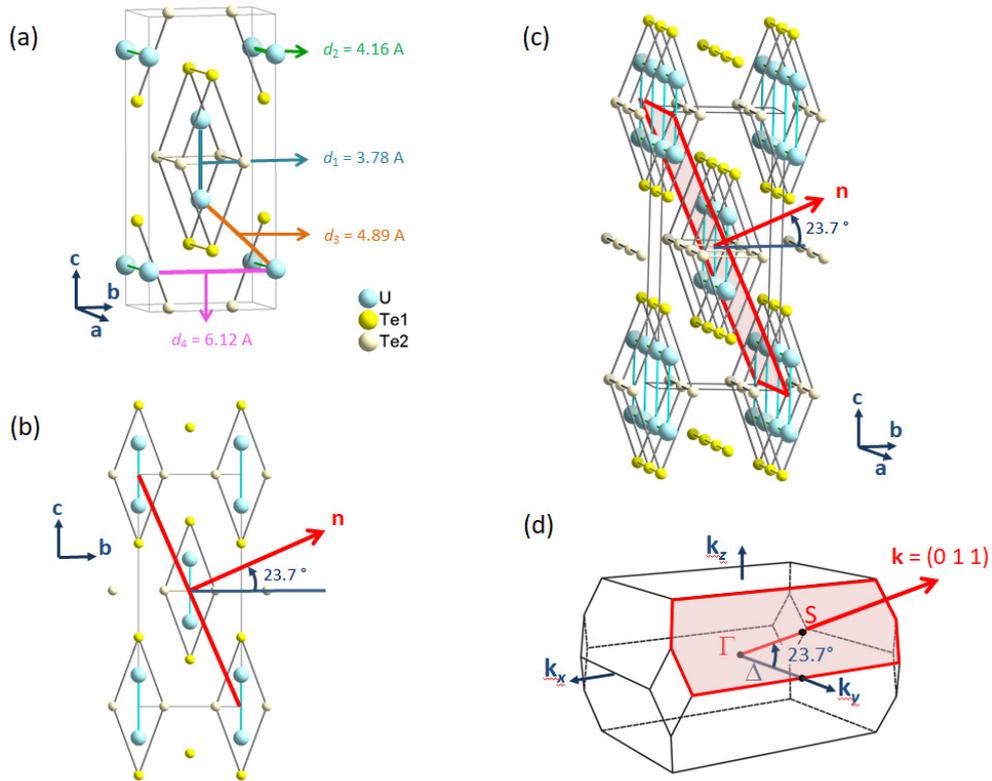

**Supplementary Figure 8. Crystal structure and Brillouin zone of UTe$_2$.** (a) Elementary unit cell and identification of the four smallest U-U distances, (b) Projection of the lattice structure in the (**b**,**c**) plane, (c) Extended crystal structure emphasizing the network of two-leg ladders, and (d) Brillouin zone of UTe$_2$. The vector **n** normal to a family of reticular (and cleaving) planes of Miller indices (0 1 1), with an angle $\theta = $ (**b**,**n**) = 23.7 °, is indicated. These reticular planes are characteristic of the ladder structure. In the reciprocal space, the corresponding wavevector **k** = (0 1 1) is perpendicular to two planes of the Brillouin zone boundary.